\documentclass{article}
\usepackage{spconf,amsmath,graphicx,balance}
\usepackage{makecell}
\usepackage{multirow}
\usepackage{color}
\usepackage{hyperref}
\usepackage{url}
\usepackage{marginnote}
\usepackage{graphicx}
\usepackage{float}
\usepackage{soul}

\title{DISCRIMINATE NATURAL VERSUS LOUDSPEAKER EMITTED SPEECH}

\name{Thanh-Ha Le$^{1,2}$, Philippe Gilberton$^1$ and Ngoc Q.~K.~Duong$^1$}
\address{$^1$Technicolor Research \& Innovation, France \\
$^2$ Eurecom, France\\
leth@eurecom.fr, \{philippe.gilberton, quang-khanh-ngoc.duong\}@technicolor.com}

\begin{document}
\maketitle

\begin{abstract}
In this work, we address a novel, but potentially emerging, problem of discriminating the natural human voices and those played back by any kind of audio devices in the context of interactions with in-house voice user interface. The tackled problem may find relevant applications in  
(1) the far-field voice interactions of vocal interfaces such as Amazon Echo, Google Home, Facebook Portal, \emph{etc}, and (2) the replay spoofing attack detection. The detection of loudspeaker emitted speech will help avoiding false wake-ups or unintended interactions with the devices in the first application, while eliminating attacks involve the replay of recordings collected from enrolled speakers in the second one. At first we collect a real-world dataset under well-controlled conditions containing two classes: recorded speeches directly spoken by numerous people (considered as the \emph{natural} speech), and recorded speeches played back from various loudspeakers (considered as the loudspeaker emitted speech). Then from this dataset, we build prediction models based on Deep Neural Network (DNN) for which different combination of audio features have been considered. Experiment results confirm the feasibility of the task where the combination of audio embeddings extracted from SoundNet and VGGish network yields the classification accuracy up to about 90\%.

\end{abstract}


\section{Introduction}
\label{sec:intro}

Vocal interface devices such as the Apple Home Pod\footnote{https://www.apple.com/homepod/}, Google Home\footnote{https://store.google.com/product/google\_home}, Amazon Echo\footnote{https://www.amazon.com/echo}, and Facebook Portal\footnote{https://portal.facebook.com/} \emph{etc.,} are becoming incredibly popular at home nowadays. These smart devices use primarily the automatic speech recognition (ASR) function to recognize user's command and spoken context. When deploying in real acoustic environments, the device will actually capture speech from both the target users and those emitted from loudspeaker devices such as radio or television. To tackle this issue, existing approach was building device-directed speech detection systems \cite{device_directed1, device_directed2, device_directed3, device_directed4, device_directed5, device_directed6} so as only the detected target speech is taken into account for the interaction. To improve the detection accuracy, speaker identification function can be added where only speech from a pre-defined list of users is processed. As examples, Google Home supports maximum 6 users in one device. Amazon Echo works with multiple user accounts, but they must be switched manually by explicitly asking Alexa to do it. To support further the multi-users vocal interface, we propose to investigate in this paper another capability to distinguish between the speech directly spoken by users, we refer hereafter as \emph{natural speech}, from those emitted by loudspeaker devices. This considered problem may greatly help the vocal interface devices to avoid annoying false wake-ups, unintended interaction, or irrelevant processing of recorded sounds. 

Another important application of the considered problem is to protect automatic speaker verification (ASV) from replay spoofing attacks (RSA) where the hackers replay recordings collected from enrolled speakers in order to provoke false alarms. The RSA is known to be very challenging as recordings made with high quality hardware may be close to indistinguishable from genuine speech signals \cite{ASVspoof2017_1, ASVspoof2017_2, ASVspoof2017_3, ASVspoof2017_4}. The considered problem does the same task with the ASVspoof 2017 Challenge\cite{ASVspoof2017_4}. However, the ASVspoof 2017 dataset \cite{ASV_dataset} was not recorded under a fully well-controlled acoustic environments which didn't fit well with our in-house context. That is the reason why we just select the ASVspoof 2017 dataset as a part of our dataset.

The considered classification problem is challenging due to the variations of the natural speech from different people, the hardware devices (microphones for recording, loudspeakers to play back), and the far-field acoustic environments. Thus, in this work we first devote our effort in collecting a large-scale dataset with significant diversification of hardware devices as well as the number of recorded users in different environments. To enrich the dataset we inherit the dataset available from the ASVspoof 2017 Challenge \cite{ASV_dataset}. Then we investigate different audio features extractors like (\emph{i.e.,} Mel-frequency cepstral coefficients (MFCC), constant Q cepstral coefficients (CQCC), embeddings from SoundNet and VGGish) for the prediction model. 

The rest of the paper is organized as follows. Section \ref{sec:dataset} presents the dataset collection. Section \ref{sec:model} focuses on building DNN-based prediction models. Experimental results are discussed in Section \ref{sec:experiment}. Finally, we conclude in Section \ref{sec:conclusion}.

\section{Dataset}
\label{sec:dataset}

In this section we will first describe the inherited dataset from the ASVspoof 2017 Challenge. We then present our own data collection where acoustic environments are well-controlled. Finally we describe the data split and augmentation for training and testing the prediction model.

\subsection{Anti spoofing replay attack dataset}
\label{ssec:asv}
The ASVspoof 2017 dataset is a collection of bona fide and spoofed utterances. The bona fide are a subset of RedDots corpus \cite{reddots_corpus}. Spoofed utterances are recorded by replaying and recording bona fide utterances using different devices and acoustic environments. The replay spoofing scenario and replay configurations are particularly described in \cite{ASV_dataset}. The ASVspoof 2017 dataset is recorded in a diverse range of replay configurations including 26 playback devices, 25 recording devices and 26 acoustic environments.
In our work, we use the ASVspoof 2017 version 2.0 dataset which was released from training, development and evaluation subsets. Since ASVspoof 2017 is considered as challenging dataset, the number of samples and replay configurations in the evaluation subset is larger than training and development subsets. We repartioned the ASVspoof 2017 version 2.0 dataset into three subsets: training, validation and testing subsets which consist of 60\%, 20\% and 20\% samples, respectively.
\subsection{In-house dataset}
\label{ssec:inhouse}

We first recorded natural speech in two languages (English and French) from ten people. We then performed denoising on the recorded natural speech signals and played them back by loudspeakers to record the loudspeaker emitted samples. The details of recording configurations are summarized as follows.\\
\textbf{Acoustic environments}: we recorded the dataset in two rooms: small size of 3~m$^2$ and medium size of 15~m$^2$. In each room, ambient noise was mainly produced by air conditioning systems, but sometimes by people walking along the corridor.
\\
\textbf{Distance from speakers/loudspeakers to the microphones}: In each room, we recorded natural voices with two distances (\emph{i.e.,} one meter and three meters) from people to the microphones. Various distances in different indoor environments may diversify greatly our dataset so that it can minimize the classifier to learn on specific acoustic conditions for the two classes.
\\
\textbf{Devices}: To avoid the classifier learning on a specific device, we recorded all speech signals with two different microphones based on two different technologies (a high end one based on electret sensor and a low cost one based on MEMS sensor). For the recording of the emitted speeches, we replayed the natural speech signals by using three different loudspeakers, among which two are from a TV set as they corresponded the most to the actual source of background speech in a home environment.
\\
\textbf{Languages}: To increase the variation of language phonemes, the dataset was built using French and English languages: each person read an English paragraph, followed by a French paragraph. The duration for reading each paragraph varied from 81 seconds to 206 seconds.

\begin{figure}[t]
  \includegraphics[width=\linewidth]{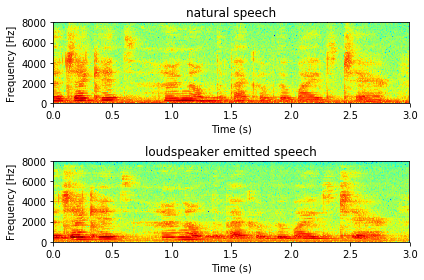}
  \caption[]{Spectrogram of the natural versus loudspeaker emitted speech of sentence 'The jacket hung on the back of the wide chair'.}
  \label{fig:spec}
\end{figure}

Overall, the dataset consisted in speech recordings upon two acoustic environments, with two different ranges between speaker/or loudspeaker and the microphone, two microphones, three loudspeakers, two languages, and ten people, resulting in a total of 480 different configurations. All samples were mono channel, recorded with a sampling rate of 16000Hz, 16 bit PCM and encapsulated into wav format. Note that, in order to avoid the machine learning to build a model that could be biased on the recording conditions between two classes (instead of learning the actual differences of spectral patterns between the natural and the loudspeaker emitted recordings), we positioned the loudspeaker at the same location as it was performed during the natural speech recording of the person. Thus the collected dataset could be more challenging than the ASVspoof 2017 dataset as each pair of samples in the two classes in the former was recorded in the identical acoustic environment. As an example, Fig. \ref{fig:spec} shows the spectrograms of a pair of recordings in such two classes, which look very similar. 
\\
\\
We segmented  the recordings into 1~s length audio segments with 50\% overlap, and partitioned all the obtained segments into 3 subsets: training, validation and testing. The number of audio samples in each set is summarized in Table \ref{table:inhouse}. Note that for a better model generalization, all these subsets were disjointed in terms of speakers.

\begin{table}[h]
\centering
\begin{tabular}{|l|l|l|l|l|l|}
\hline
\multicolumn{2}{|l|}{\#seg in training} & \multicolumn{2}{l|}{\#seg in validation} & \multicolumn{2}{l|}{\#seg in testing} \\ \hline
Natural & Emitted & Natural & Emitted & Natural & Emitted \\ \hline
5053 & 9832 & 5163 & 10263 & 8390 & 16527 \\ \hline
\end{tabular}
\caption{Number of audio samples (natural and loudspeaker emitted segments) in the training, validation, and testing set of the In-house dataset.}
\label{table:inhouse}
\end{table}

\subsection{Data organization and pre-processing}
\label{ssec:dataprocessing}

After segmenting the recordings into 1~s segments with 50\% overlap, we normalized the segments' energy to the same level. The new dataset was obtained by merging the ASVspoof 2017 dataset and the In-house dataset. We merged together the training set, validation set and testing set of these two dataset. The final number audio samples in each set of the overall dataset is shown in Table \ref{table:data}. As it can be seen, the dataset is not perfectly balanced as the number of loudpeaker emitted segments is about three times bigger than the number of natural speech segments in each subset.

\begin{table}[h]
\centering
\begin{tabular}{|l|l|l|l|l|l|}
\hline
\multicolumn{2}{|l|}{\#seg in training} & \multicolumn{2}{l|}{\#seg in validation} & \multicolumn{2}{l|}{\#seg in testing} \\ \hline
Natural & Emitted & Natural & Emitted & Natural & Emitted \\ \hline
15068 & 51186 & 8386 & 24319 & 11409 & 30270 \\ \hline
\end{tabular}
\caption{Total number of audio samples (natural and loudspeaker emitted segments) in the training, validation, and testing set of the overall dataset.}
\label{table:data}
\end{table}
\section{Prediction model}
\label{sec:model}
Prior to building the prediction model, we first extracted a range of state-of-the-art audio features for each 1~s audio samples in the dataset. The details of the extracted features will be presented in Section \ref{ssec:feature}. We then investigated the use of multi-layer perception (MLP) and support vector machine (SVM) as classifier. After using grid search to optimize the parameters for both MLP and SVM, we found that MLP gave slightly better prediction accuracy than SVM. Thus we will focus preferably on MLP that will be our choice of classifier for which we will report its results in the experiment.

\subsection{Audio feature extraction}
\label{ssec:feature}
\textbf{MFCC}: as a well-known feature extractor for automatic speech recognition, in our investigation we extracted 12 MFCC coefficients for each overlapping audio frame of 25 miliseconds (hope size is 10 miliseconds). As a result, we obtained a 1212-dimensional MFCC feature for each 1~s audio sample in the dataset. To make the feature dimension comparable to VGGish and SoundNet features, we further performed a Principal Component Analysis (PCA) to reduce the MFCC feature dimension to 128. The 128-dimensional feature may be more relevant for training the model given the moderate size of the dataset. In the experiment, we will report prediction results obtained with both the 1212-dimensional and 128-dimensional MFCC features.
\\
\textbf{CQCC}: this feature uses constant Q transform to offer greater frequency resolution in lower frequencies and greater time resolution in higher frequencies. As loudspeaker devices often make more signal distortion at low and high frequency ranges, we expect that the CQCC can distinguish such differences between the natural versus loudspeaker emitted speech, and thus potentially improve the prediction result. Note that CQCC was also used as main feature in the baseline system of the ASVspoof 2017 challenge \cite{CQCC_feature}. In our experiment, the constant Q transform is applied with a maximum frequency of 8000Hz, the minimum frequency is set to 15Hz, the number of bins per octaves is 96 and the sampling period is 16. Each 1~s audio sample in the dataset results in a 1404-dimensional CQCC feature. Similarly to the MFCC, we further used PCA to reduce the feature dimension to 128 to be balance with other DNN-based features and to be more relevant with the actual size of the dataset.
\\
\textbf{SoundNet feature}: SoundNet was designed to learn rich natural sound representations from large amounts of unlabeled sound data collected in the wild \cite{SoundNet}. It was targeted to transfer discriminative visual knowledge from convolutional neural network (CNN)-based visual recognition models into the sound modality using 2,000,000 unlabeled videos. Soundnet’s embedding features were reported to obtain the state-of-the-art result for acoustic scene classification task on three benchmark datasets: DCASE Challenge \cite{DCASEpaper, dcase2017}, ESC-50 and ESC-10 \cite{ESCpaper}. In our experiment, we used the pre-trained SoundNet based on 8 convolutional layers and we extracted embeddings from different layers for testing: conv4, conv5, conv6, and conv7. The best classification result (reported in the experiment section) was obtained with the 512-dimensional embedding at the layer conv5.
\\
\textbf{VGGish feature}: Released recently by Google\footnote{https://github.com/tensorflow/models/tree/master/research/audioset}, VGGish is a deep convolutional neural network trained on Audioset dataset \cite{VGGish}. Audioset is a dataset of over two million human-labeled 10-second YouTube video soundtracks, with labels of more than 600 audio event classes \cite{AudioSet}. VGGish is a variant of the well-known VGG model designed for image classification. The input size was 96x64 for log mel spectrogram audio inputs. The last group of convolutional and maxpool layers is dropped, so that VIGGish has four groups of convolution/maxpool layers instead of five as the VGG11. At the end of the network, a 128-wide fully connected layer is used and this layer acts as a compact embedding layer. In our experiment, we extract this 128-dimensional VGGish embedding for each 1-s audio sample in the dataset.
\\
\textbf{Combined features}: Since SoundNet was trained with unlabeled data and achieved state-of-the-art results in acoustic scene detection benchmark datasets, and VGGish was trained with labeled audio object classes, we concatenated embeddings extracted from those two neural networks with expectation that the new features can benefit from representing both information about audio scenes and audio objects. We also investigated the combination (by concatenation also) of all above mentioned features after balancing their dimension (\emph{i.e.,} to 128 for MFCC and CQCC).

\subsection{Model training}
\label{ssec:training}

We tested two classification models: SVM and MLP for each type of features described in the Section \ref{ssec:feature}. During the training, we  varied the model parameters with grid search, \emph{i.e.,}  kernel types and their corresponding hyper-parameters for SVM;  number of layers, number of neurons in each layer, batch size, step size, optimizer, \emph{etc.,} for the MLP. We also performed the 5-fold cross validation to choose the best performing model in term of the classification accuracy.

\section{RESULT}

The best results offered by each type of audio feature were obtained by averaging the corresponding classification accuracy in each fold and shown in Table \ref{table:results}. Note that we do not show the results given by the SVM as in all cases MLP performs better. We also report the results of a Gaussian mixture model (GMM) classifier that was trained with CQCC feature as baseline model of the ASVspoof 2017 Challenge\cite{ASVspoof2017_4}.

Compared to the baseline model of ASVspoof 2017 Challenge, all MLP classifiers performed better. As expected, 128 dimensional CQCC outperformed MFCC in the validation and test sets as CQCC was designed to better capture acoustic variations at low and high frequencies. On the other hand, DNN-based features SoundNet and VGGish performed better than the signal processing based features MFCC and CQCC as DNN trained on numerous data with different recording quality may capture richer information about audio signal, especially at low and high frequency ranges where the difference between the two sound classes may be more visible. This was also in line with current observations in the acoustic scene classification\footnote{http://dcase.community/challenge2018/task-acoustic-scene-classification} \cite{SoundNet} and event detection tasks\footnote{http://www.cs.tut.fi/sgn/arg/dcase2017/challenge/task-sound-event-detection-in-real-life-audio}. VGGish feature, which was released recently by Google, offers the best testing accuracy of 88.04\%.  Overall the combination of VGGish and SoundNet features offer the best classification performance for both the training (90.04\%), validation  (88.54\%), and testing  (90.43\%) sets. We also observed that a further combination of VGGish and SoundNet with MFCC and CQCC did not bring benefit as there might be acoustic redundancy in such combination. 

The best MLP model was trained with combination of VGGish and SoundNet features has one fully connected layer with 100 neurons and softmax is used in the last layer. The learning rate is 0.00005, batch size is 5000 and the optimizer is Adam.

\label{sec:experiment}
\begin{table}[h]
\centering
\begin{tabular}{|l|c|c|c|c|}
\hline
\multirow{2}{*}{Audio features} & \multirow{2}{*}{\#dim.} & \multicolumn{3}{c|}{Accuracy \%} \\ \cline{3-5} 
 &  & Training & Validation & Testing \\ \hline
MFCC & 1212 & 83.10 & 83.84 & 84.05 \\ \hline
MFCC & 128 & 89.84 & 84.93 & 84.79 \\ \hline
CQCC & 1404 & 87.58 & 84.64 & 84.95 \\ \hline
CQCC & 128 & 86.59 & 85.53 & 85.29 \\ \hline
SoundNet & 512 & 89.45 & 86.00 & 86.87 \\ \hline
VGGish & 128 & 88.58 & 86.05 & 88.04 \\ \hline
\textbf{VGGish+SoundNet} & \textbf{640} & \textbf{90.40} & \textbf{88.54} & \textbf{90.43} \\ \hline
\begin{tabular}[c]{@{}l@{}}VGGish+SoundNet\\ + CQCC+MFCC\end{tabular} & 512 & 89.77 & 86.36 & 88.68 \\ \hline
\begin{tabular}[c]{@{}l@{}}Baseline model\\ of ASVspoof 2017\end{tabular} & 2223 & 76.31 & 76.30 & 72.63 \\ \hline
\end{tabular}
\caption{Classification results on the training, validation, and test set obtained by MLP classifier with the use of different audio features.}
\label{table:results}
\end{table}

\section{CONCLUSION}
\label{sec:conclusion}

In this work, we discussed a novel problem of detecting whether a speech recording is \emph{natural}, \emph{i.e.,} directly spoken by a person, or emitted from a loudspeaker device. Reliable detection model can find usage in emerging vocal interfaces and anti-spoofing application. We designed a large-scale dataset under well-controlled conditions to better fit with home environment and to allow us to train DNN-based prediction models. The best system with 90.43\% test accuracy was obtained by using the combination of VGGish and SoundNet embeddings as audio feature extractor and MLP as classifier. This result confirms the feasibility of the task despite such a challenging classification problem. Future work may be devoted in enlarging the dataset with more users and more variety of loudspeaker/microphone devices to confirm the results. Fine-tuning VGGish or SoundNet so as the prediction system is learned end-to-end directly from input audio recordings is also a potential way to improve the prediction accuracy, especially with the enlarged dataset. Finally, a more analysis about what is the essence for discriminating the two classes of speech signals is planned.

\bibliographystyle{IEEEbib}
\balance
\bibliography{refs}
\end{document}